\documentstyle [12pt] {article}
\title{SINGLE HORIZON BLACK HOLE "LASER" AND A SOLUTION OF THE
INFORMATION LOSS PARADOX}

\author{Vladan Pankovi\'c$^{\ast,\sharp}$, Rade Glavatovi\'c$^\diamond$,
Simo Ciganovi\'c$^\sharp$, \\ Du\v{s}an - Harper Petkovi\'c $^\sharp$ , Lovro - Loka Martinovi\'c $^\sharp$\\
$^\ast$Department of Physics, Faculty of Sciences, 21000 Novi
Sad,\\ Trg Dositeja Obradovi\'ca 4. , Serbia, vdpan@neobee.net \\
$^\sharp$Gimnazija, 22320 Indjija, Trg Slobode 2a, Serbia\\
$^\diamond$ Military-Medical Academy, 11000 Belgrade, Crnotravska
17., Serbia \\}

\date {}
\begin{document}
\maketitle

\vspace{0.5cm}


 \vspace{0.5cm}

\begin {abstract}
In this work we show that single horizon black hole behaves as a
"laser". It is in many aspects conceptually analogous to Corley
and Jacobson work on the two horizon black hole "laser". We
started by proposition that circumference of the black hole
horizon holds the natural (integer) quantum number of
corresponding reduced Compton's wave length of some boson systems
in great canonical ensemble. For macroscopic black hole ground
state is practically totally occupied while other states are
practically totally unoccupied which is a typical Bose
condensation. Number of the systems in this condensate represents
black hole entropy. For microscopic black hole few lowest energy
levels are occupied with almost equivalent population (with
negative chemical potential) while all other energy states (with
positive chemical potential) are practically unoccupied. It
implies that here not only spontaneous but also stimulated
emission of radiation comparable with spontaneous emission occurs.
By Hawking evaporation any macroscopic black hole turns out in a
microscopic black hole that yields, in a significant degree,
coherent stimulated emission of the radiation. It implies that by
total black hole evaporation there is no decoherence, i.e.
information loss. Finally, a mass duality characteristic for
suggested black hole model corresponding to string T-duality is
discussed.
\end {abstract}

\newpage

 "On Volodja Vysotsky I decided to write a ballad \\
On another solider, from horizon, irreversible coming no back.\\
One can say: he was a great sinner without lucky light\\
But nature does not know a singular singer without sin\\
(nor without innocence track).\\
…\\
On Volodja Vysotsky I tended to make a poem in voice law\\
but my arm was decoherently shaking and verse avoided my pen.\\
Black crane/swon in the black hole tended to go\\
but correlated coherent white crane/swon is in the white universe again."\\
\\
\begin{flushright}Bulat Okudzava, "Poem on Vladimir Vysotsky"\\
(translated in free style by Vladan Pankovi\'c and Darko Kapor)\\
\end{flushright}

\vspace {0.5cm}

\section {Introduction}

As it is known Corley and Jacobson [1] have shown that a two
horizon boson black hole can behave as a "laser" that amplifies
Hawking radiation. Precisely, Corley and Jacobson have shown that:
"High energy frequency spectrum of the Hawking radiation from a
single black hole horizon, whether the dispersion entails
subluminal or superluminal group velocities. … in presence of an
inner horizon as well as an outer horizon the superluminal case
differs dramatically however. The negative energy partners of
Hawking quanta return to the outer horizon and stimulate more
Hawking radiation if the field is bosonic or suppress it if the
field is fermionic." [1]

Recently, Leonhard and Philbin [2] provided some new numerical
result that refers on the Corley-Jacobson predictions whose
detailed analytic theory is very complicated. "The production of
Hawking radiation by a single horizon is not dependent on the
high-frequency dispersion relation of the radiated field. When
there are two horizons, however, Corley and Jacobson have shown
that superluminal dispersion leads to an amplification of the
particle production in the case of bosons. The analytic theory of
this "black hole laser" process is quite complicated, so we
provide some numerical results in hope of aiding understanding of
this interesting phenomenon." [2]

In our previous works [3]-[5], we reproduced and determined in the
simplest way three well-known [6]-[12], most important
thermodynamical characteristics (Bekenstein-Hawking entropy,
Bekenstein quantization of the entropy or horizon surface area and
Hawking temperature) of Kerr-Newman (Schwarzschild, Kerr,
Reissner-Nordstr$\ddot {o}$m) black hole. We started  physically
by assumption that circumference of black hole horizon holds the
natural (integer) number of corresponding reduced Compton's wave
length so that given numbers determine corresponding energy
levels. Mathematically, we use, practically, only simple algebraic
equations. (It is conceptually similar to Bohr's quantization
postulate in Bohr's atomic model interpreted by de Broglie
relation.)

Our simple, "macroscopic" predictions are in an excellent
agreement not only with standard ("mesoscopic") black hole
thermodynamics [6]-[12] but also with Copeland and Lahiri work
[13]. Namely, Copeland and Lahiri, starting from a possible
"microscopic", i.e. string theory, demonstrated that
thermodynamical characteristics of (Schwarzschild) black hole can
be obtained by a standing waves corresponding to small
oscillations on a circular loop with radius equivalent to
(Schwarzschild) horizon radius. It indicates that our results even
simply have not only a formal meaning.

In this work, generalizing our previous results, we shall show
that single horizon black hole, i.e. Schwarzschild black hole,
behaves as a "laser". It is not contradictory but, conceptually,
in many aspects analogous to Corley and Jacobson theory. We again
shall start by proposition that circumference of the black hole
horizon holds the natural (integer) quantum number of
corresponding reduced Compton's wave length of some bosonic
systems. More precisely, we shall suggest a simplified statistical
model of the black hole according to which black hole consists of
two parts, first one - a large system, i.e. a "nucleus" or
statistical reservoir, and second one - a small system, i.e. a
"shell" or a  great canonical ensemble of mentioned bosonic
systems.

For macroscopic black hole (with mass larger than Planck mass)
chemical potential is always positive that expresses diminishing
of the energy of ensemble by reservoir. Also ground state is
practically totally occupied while other states are practically
totally unoccupied that is a typical Bose condensation. Number of
the systems in this condensate (multiplied by Boltzmann constant)
represents black hole entropy which yields a simple explanation of
the black hole entropy. Obviously, for single horizon macroscopic
black hole spontaneous Hawking emission of the radiation
represents unique possible form of the emission. Stimulated
emission of the radiation is here impossible. It is in full
agreement with Corley and Jacobson theory.

For microscopic black hole (with mass smaller than Planck mass)
few lowest energy levels (ground state and some its neighboring
states) are occupied with almost equivalent population and with
negative chemical potential. It expresses energy stimulation of
the ensemble by reservoir. All other energy states, with positive
chemical potential, are practically unoccupied. It implies that
now not only spontaneous but also stimulated emission of radiation
comparable with spontaneous emission occurs. Simply speaking we
shall show that single horizon microscopic black hole represents
really a "laser". It is not hard to see that presented situation
is conceptually analogous to Corley and Jacobson theory (e.g.
negative chemical potential corresponds conceptually to return of
the negative energy partners of Hawking quanta, etc.).

It can be observed that in distinction from spontaneous, Hawking
emission of the radiation by black hole that is decoherent,
stimulated emission of the radiation by black hole is coherent. By
Hawking evaporation any macroscopic black hole turns out in a
microscopic black hole that yields partially coherent stimulated
emission of the radiation. It implies, as it will be shown, that
by total black hole evaporation there is no decoherence, i.e.
information loss. In this way stimulated emission of the single
horizon black hole can solve old problem of the information loss
[14]-[17].

Finally, a mass duality characteristic for suggested black hole
model corresponding to string T-duality will be discussed.

\section {A simple determination of the black hole thermodynamical characteristics
and a simple model of the black hole}

Firstly, we shall shortly repeat our previous results [3]-[5].

Analyze a Schwarzschild's black hole with mass $M$ and
Schwarzschild's radius
\begin {equation}
      R = \frac {2GM}{c^{2}}
\end {equation}
where $G$ represents Newtonian gravitational constant and $c$ -
speed of light.

Introduce the following condition
\begin {equation}
      m_{n}c R = n\frac {\hbar}{2p}  \hspace{1cm}   {\rm for}  \hspace{1cm}   n = 1,
      2,...
\end {equation}
,where $\hbar$ represents reduced Planck constant, what implies
\begin {equation}
      2\pi R = n \frac {\hbar }{m_{n}c} = n \lambda_{rn} \hspace{1cm}   {\rm for}  \hspace{1cm}    n = 1, 2,...            .
\end {equation}
Here $2\pi R $ represents the circumference of the black hole
horizon while
\begin {equation}
        \lambda_{rn}=\frac {\hbar }{m_{n}c}
\end {equation}
is $n$-th reduced Compton wavelength of a quantum system with mass
$ m_{n}$ captured at the black hole horizon surface for $n = 1, 2,
…$ . Expression (3) simply means that {\it circumference of the
black hole horizon holds exactly} $n$ {\it corresponding} $n$-{\it
th reduced Compton wave lengths of a quantum system with mass} $
m_{n}$ {\it captured at the black hole horizon surface}, for $n =
1, 2, …$ . Obviously, it is essentially analogous to well-known
Bohr's angular momentum quantization postulate interpreted via de
Broglie relation. However, there is a principal difference with
respect to Bohr's atomic model. Namely, in Bohr's atomic model
different quantum numbers $n = 1, 2, …$ , correspond to different
circular orbits (with circumferences proportional to $n^{2} =
1^{2},  2^{2}, …$). Here any quantum number $n = 1, 2, …$
corresponds to the same circular orbit (with circumference $2\pi R
$).

According to (2) it follows
\begin {equation}
       m_{n} = n\frac {\hbar }{2\pi cR}= n \frac {\hbar c}{4\pi GM }\equiv n m_{1}
        \hspace{1cm}   {\rm for}  \hspace{1cm} n = 1, 2,...
\end {equation}
where minimal system mass equals
\begin {equation}
       m_{1} = \frac {\hbar c}{4\pi GM } =\frac {1}{4\pi} \frac {M_{P}}{M}M_{P} = \frac {1}{4\pi} \frac {MP}{M}
\end {equation}
and where $MP=(\hbar c /G)^{\frac {1}{2}}$ is the Planck mass.
Obviously, $ m_{1}$ depends of $M$ so that $ m_{1}$ decreases when
$M$ increases and vice versa. For a "macroscopic" black hole, i.e.
for $M \gg M_{P}$ it follows $m_{1}\ll M){P}\ll M$. But, for a
"microscopic" black hole, i.e. for $M \leq M_{P}$ it follows
$m_{1}\geq M_{P}\geq M$.

Define now the following
\begin {equation}
       \sigma = \frac {M}{ m_{1}} = 4\pi \frac {GM^{2}}{\hbar c}
\end {equation}
that, after multiplication by Boltzmann constant $k_{B}$, yields
\begin {equation}
       S = k_{B}\sigma  = 4\pi k_{B} \frac {GM^{2}}{\hbar c} = k_{B}\frac {c^{3}}{4G\hbar}A = k_{B}\frac {A}{4L^{2}_{P}}
\end {equation}
where
\begin {equation}
       A = 4\pi R^{2}
\end {equation}
represents black hole surface area, while $L_{P}=(G{
\hbar}/c^{3})^{\frac {1}{2}}$ represents Planck length. Obviously,
(8) represents Bekenstein-Hawking entropy of the black hole.

Differentiation of (8) yields
\begin {equation}
      dS = k_{B}d\sigma = 8\pi k_{B}\frac {GM}{\hbar c}dM= 8\pi k_{B}\frac {GM}{\hbar c^{3}}d(Mc^{2}) =
      8\pi k_{B}\frac {GM}{\hbar c^{3}}dE
\end {equation}
where
\begin {equation}
      E=Mc^{2}
\end {equation}
represents the black hole total energy. Expression (10), according
to first thermodynamical law, implies that term
\begin {equation}
      T =\frac  {\hbar c^{3}}{8\pi k_{B}GM}
\end {equation}
represents the black hole temperature. Evidently, this temperature
is identical to Hawking black hole temperature.

Further, according to (8)-(10) it follows
\begin {equation}
       dA = (32\pi \frac {G^{2}}{c^{3}}) M dM
\end {equation}
or, in a corresponding finite difference form
\begin {equation}
       \Delta A = (32\pi \frac {G^{2}}{c^{3}})  M \Delta M  \hspace{1cm}   {\rm for}  \hspace{1cm}   \Delta M \ll M .
\end {equation}
Now, assume
\begin {equation}
       \Delta M= n m_{1} \hspace{1cm}   {\rm for}  \hspace{1cm} n = 1,
       2,...
\end {equation}
which, according to (6), after substituting in (14), yields
\begin {equation}
       \Delta A = n 8L^{2}_{P}  \hspace{1cm}   {\rm for}  \hspace{1cm} n = 1, 2,...           .
\end {equation}
Obviously, expression (16) represents Bekenstein quantization of
the black hole horizon surface area.

Now, according to presented results, a simple, effective model of
the black hole can be suggested.

According to this model black hole can be approximately considered
as a quantum super-system, $L+s$, which holds two quantum
sub-systems.

First one is a "large" quantum sub-system ("nucleus"), $L$, inside
horizon with mass $M^{L}= M$ and energy $E^{L}=E$ described by
pure quantum state $|E^{L}>=|E>$.

Second one is "small" quantum sub-system, $s$, on the horizon
surface described by a statistical mixture of the quantum states
$|E^{s}_{n}>$ for $n = 1, 2, …$ corresponding to masses $m_{n}$
for $n = 1, 2, …$ (2), and energies
\begin {equation}
      E_{n} = m_{n}c^{2} = n \frac {\hbar c^{3}}{4\pi MG} =
      n \frac {1}{4\pi}\frac {M_{P}}{M}E_{P} = n E_{1} \hspace{1cm}   {\rm for}  \hspace{1cm} n = 1,
      2,...
\end {equation}
where $E_{P}= M_{P}c^{2}$ is Planck energy and
\begin {equation}
      E_{1}= m_{1} c^{2}  =  \frac {\hbar c^{3}}{4\pi MG}      .
\end {equation}
For a macroscopic black hole, i.e. for $M \gg M_{P}$ it follows
$E_{1} \ll  E_{P}$. Obviously, (17) corresponds to energy spectrum
of a linear harmonic oscillator. In any of quantum states
$|E^{s}_{n}>$ for $n = 1, 2,...$, $s$ occupies circumference of
the horizon.

Finally, it will be supposed that, in the first approximation,
dynamical interaction between $s$ and $L$ can be effectively
reduced at the dynamical evolution and statistical distribution
(that will be discussed later) of the quantum states of $s$
without changing of the quantum states of $L$.

Thus, in our simple model black hole holds a form in some degree
similar to form of Bohr's atom where electron propagates along
circular orbits around atomic nucleus. (But, as it has been
already pointed out, there is a principal difference in respect to
Bohr's atomic model.)

In this way we have reproduced, i.e. determined exactly in a
mathematically and physically simple way, three most important
characteristics of Schwarzschild's black hole thermodynamics,
Bekenstein-Hawking entropy (8), Hawking temperature (12), and
Bekenstein quantization of the horizon surface area (16).

\section {Single horizon black hole "laser"}

Now we shall demonstrate that suggested model of the black hole
admits consistent theoretical description of the stimulated
emission of the radiation by black hole.

Suppose that s has been initially in some lower quantum state
$|E^{s}_{k}>$ and that higher quantum state $|E^{s}_{n}>$ , for $k
< n$, has been initially "empty", i.e. "unoccupied". (Concrete
meaning of the words "empty", i.e. "unoccupied" or "occupied"
depend of the concrete, fermionic or bosonic, nature of $s$.)

Suppose that black hole absorbs a quantum system with total energy
\begin {equation}
        \epsilon_{nk} = E_{n} - E_{k}                              .
\end {equation}
According to no-hair theorem and supposition that given black hole
is Schwarzschild given energy characterizes practically completely
given quantum system. Then given absorption can be presented as
transition of $s$ in the final higher quantum state $|E^{s}_{n}>$
.

Unitary character of the quantum dynamics that describe presented
absorption admits that opposite situation, i.e. stimulated
emission of the radiation by black hole can be described in the
following way.

Suppose that $s$ has been initially in some higher quantum state
$|E^{s}_{n}>$  and that lower quantum state $|E^{s}_{k}>$ , for $k
< n$, has been initially "empty", i.e. "unoccupied". Suppose that
an external quantum system, $SE$, with total energy
$\epsilon_{nk}$ (19), propagates nearly but without black hole
horizon. $SE$ can be, also, a quant of Hawking radiation emited in
the previous time moment by black hole. Then, without any
absorption of $SE$ by black hole and under resonant dynamical
influence of $SE$ at black hole, $s$ turns out in the final, lower
quantum state $|E^{s}_{k}>$.  So, finally, there are two quantum
systems with the same energy $\epsilon_{nk}$  (19) that propagate
outside black hole. In this sense, roughly speaking, black hole
can be considered as a "laser".

But, of course, black hole can be considered as a "laser" if and
only if the inverse population of $s$ can be realized at least in
some degree, i.e. at least in case when a lower and higher energy
level have almost equivalent populations. It needs a more detailed
analysis of the statistical characteristics of $s$ in respect to
our previous works [3]-[5].

Suppose that black hole, precisely $s$, can be considered as a
great canonical statistical ensemble of the ideal, non-interacting
Bose-Einstein quantum systems. In this case $L$ can be considered
as reservoir.

Then, as it is well-known, statistical sum can be presented by
expression
\begin {equation}
  Z = \sum_{n,N(n)=0}\exp[-N(n)\frac {E_{n}-\mu}{k_{B}T}] = \sum_{n=0} Z_{n}      .
\end {equation}
Here $\mu$ represents the chemical potential, $N(n)$ - number of
the systems in quantum state of the individual system
$|E^{s}_{n}>$ with energy $E_{n}$, for $n=1, 2, …$, while
\begin {equation}
   Z_{n} = \sum_{N(n)=0}\exp[-N(n) \frac {E_{n}-\mu}{k_{B}T}]
\end {equation}
represents the partial statistical sum for $n=1, 2, …$ .

Also, corresponding partial probabilities, i.e. statistical
weights of the event that $N(n)$ individual systems have energy
$E_{n}$
\begin {equation}
   w_{N(n)} = Z^{-1}_{n}\exp[-N(n) \frac {E_{n}-\mu}{k_{B}T}] \hspace{1cm}   {\rm for}  \hspace{1cm} N(n)=0, 1,
   2,...
\end {equation}
can be defined, for $n=1, 2, …$ .

Then, as it is well-known, it follows quite generally
\begin {equation}
   Z_{n}= (1 -  \exp[-\frac {E_{n}-\mu}{k_{B}T}])^{-1} \hspace{1cm}   {\rm for}  \hspace{1cm} n=1,
   2,...
\end {equation}
\begin {equation}
  <N>_{n} = (\exp[\frac {E_{n}-\mu}{k_{B}T}] - 1) ^{-1} \hspace{1cm}   {\rm for}  \hspace{1cm} n=1,
  2,...
\end {equation}
\begin {equation}
   N_{TOT}= \sum_{n=1}<N>_{n} \hspace{1cm}   {\rm for}  \hspace{1cm} n=1,
   2,...
\end {equation}
\begin {equation}
   E _{TOT} = \sum_{n=1}<N>_{n} E_{n} \hspace{1cm}   {\rm for}  \hspace{1cm} n=1,
   2,...
\end {equation}
where $<N>_{n}$ represents the average number of the quantum
systems in quantum state $|E^{s}_{n}>$ for $n=1, 2, …,$ $N _{TOT}$
- statistically averaged total number of the quantum systems, and
$E _{TOT}$ - statistically averaged total energy of the quantum
systems.

It can be supposed
\begin {equation}
   E _{TOT} = E
\end {equation}
so that, according to (17), (18) , it follows
\begin {equation}
   M = \sum_{n=1} <N>_{n}m_{n}  \hspace{1cm}   {\rm for}  \hspace{1cm} n=1,
   2,...
\end {equation}
and further
\begin {equation}
   \frac {M}{m_{1}} = \sum_{n=1} <N>_{n}n  \hspace{1cm}   {\rm for}  \hspace{1cm} n=1, 2,...            .
\end {equation}
For a macroscopic black hole, i.e. for
\begin {equation}
  M \gg M_{P}\gg m_{1}
\end {equation}
or
\begin {equation}
   \sigma \gg 1
\end {equation}
left hand of (29) represents a large number. It implies that right
hand of (29) can be approximated by first term in the sum. Namely,
according to (17), (24), it follows
\begin {equation}
    <N>_{n}n \simeq n \exp [-2n] \simeq 0
    \hspace{1cm}   {\rm for}\hspace{0.2cm} {\rm large} \hspace{0.2cm} n \hspace{1cm} i.e. \hspace{0.2cm}n \gg 1  .
\end {equation}
For this reason practically all terms, except few first, in the
right hand of (29) can be approximately neglected. Given
non-neglectable few first terms can be determined in the following
way. Linear Taylor expansion of $<N>_{n}$ (24) yields
\begin {equation}
  <N>_{n}\simeq \frac {k_{B}T}{E_{n}-\mu}
  \hspace{1cm}   {\rm for} \hspace{0.2cm} {\rm small} \hspace{0.2cm}n  \hspace{1cm}  {\rm i.e. \hspace{0.2cm}for} \hspace{0.2cm} n \sim1,
\end {equation}
so that
\begin {equation}
  <N>_{n}n \simeq n \frac {k_{B}T}{E_{n}-\mu}
  \hspace{1cm}   {\rm for} \hspace{0.2cm} {\rm small} \hspace{0.2cm}n \hspace{1cm} {\rm i.e. \hspace{0.2cm}for} \hspace{0.2cm} n \sim1 .
\end {equation}

In the simplest approximation, i.e. by reduction of (29) in the
first term, i.e. for $n=1$, according to (34), (29) turns out
approximately in
\begin {equation}
  \frac {M}{m_{1}}\simeq \frac {k_{B}T}{E_{1}-\mu}   .
\end {equation}
It, according to (6), (8), (17), (18), yields
\begin {equation}
  \mu \simeq E_{1}-(\frac {m_{1}}{M})(k_{B}T ) = E_{1}(1 - \frac {1}{2\sigma}) =
  \frac {1}{4\pi}\frac {M_{P}}{M}(M_{P}c^{2})[1 - \frac {1}{8\pi}(\frac { M_{P}}{M})^{2}] .
\end {equation}
Obviously, it means, according to (31), that chemical potential µ
represents a $M$ dependent function. It, metaphorically speaking,
represents a scaling. Also, chemical potential is positive, which
means that here statistical ensemble energy is diminished by
reservoir.

Now, we shall introduce (36) in (33), that yields
\begin {equation}
 <N>_{1}\simeq \frac {M}{m_{1}} = \sigma = \frac {S}{k_{B}}\gg 1
\end {equation}
\begin {equation}
<N>_{2}\simeq \frac {1}{2}\ll <N>_{1}
\end {equation}
which means that given approximation is satisfactory.

So for a macroscopic black hole, i.e. when condition (30) or (31)
is satisfied, expressions (37),(38) point out that practically all
quantum systems from statistical ensemble, occupy ground quantum
state $|E^{s}_{1}>$ . It can be effectively treated as a
Bose-Einstein condensation. Also, it points out unambiguously that
here inverse population cannot exist, even not approximately. In
other words, for a macroscopic black hole for one horizon there is
no stimulated emission of the radiation.

Also, it can be pointed out that for a macroscopic black hole,
i.e. when condition (30) or (31) is satisfied, black hole entropy
(divided by Boltzmann constant) according to (35), is practically
equivalent to number of the quantum systems in the ground state.
It represents a simple and clear (non-obscure) interpretation of
the black hole entropy.

Consider now a microscopic black hole for which condition
\begin {equation}
   M_{1} \geq M_{P}\geq M
\end {equation}
or
\begin {equation}
   \sigma = \frac {S}{k_{B}}\leq 1
\end {equation}
is satisfied.

Now, left hand of (29) represents a small number (smaller than 1)
so that any term on the right hand of (29) must be small number
(smaller than 1) too. Here condition, i.e. expression (32), is
satisfied too. For this reason all terms in the right hand of (29)
corresponding to large $n$ must be again neglected. Also, right
hand of (29) must be again approximated by few first terms. It, in
the simplest approximation, i.e. by reduction of (29) in the first
term, i.e. for $n=1$, yields again (35) and (36). Meanwhile,
since, according to (40), $\sigma$ is small (smaller than 1),
$E_{1}-\mu$ must be large (significantly larger than $k_{B}T$ ).
Especially, according to (36), for
\begin {equation}
   \sigma < \frac {1}{2}
\end {equation}
or for
\begin {equation}
   M < \frac {M_{P}}{(8\pi)^{\frac {1}{2}}}
\end {equation}
chemical potential $\mu$ becomes negative. It, roughly speaking,
means that reservoir amplifies energy of the statistical ensemble.

Now, we shall introduce (36) in (33), that, for $n=1,2 $ yields
\begin {equation}
    <N>_{1} = \sigma =  \frac {S}{k_{B}} < 1
\end {equation}
\begin {equation}
    <N>_{2} \simeq \frac {1}{2 + 1/ \sigma} =
    \frac {\sigma}{2 \sigma + 1}\leq \sigma  \hspace{1cm}   {\rm for}  \hspace{1cm} \sigma \ll \frac {1}{2}  .
\end {equation}

So for a microscopic black hole, i.e. when condition (41) or (42)
is satisfied, expressions (43),(44) point out that quantum systems
from statistical ensemble, occupy practically equivalently ground
quantum state $|E^{s}_{1}>$ and first excited state $|E^{s}_{2}>$.
It points out unambiguously that here in some degree inverse
population exists, or, precisely, that here both emissions,
spontaneous and stimulated, are almost equivalently probable. In
other words, for a macroscopic bosonic black hole with one horizon
there is a significant probability for the stimulated emission of
the radiation. In this way a microscopic bosonic black hole can be
(partially) considered as a "laser".

Suppose now that black hole, precisely s, can be considered as a
great canonical statistical ensemble of the ideal, non-interacting
Fermi-Dirac quantum systems.

In this case, as it is well-known, it follows
\begin {equation}
   Z_{n} = (1 +  \exp[-\frac {E_{n}-\mu}{k_{B}T}])  \hspace{1cm}   {\rm for}  \hspace{1cm} n=1,
   2,...
\end {equation}
\begin {equation}
   <N>_{n} = (\exp[-\frac {E_{n}-\mu}{k_{B}T}] + 1)^{-1}  \hspace{1cm}   {\rm for}  \hspace{1cm} n=1,
   2,...
\end {equation}
\begin {equation}
   N_{TOT} = \sum_{n=1}<N>_{n} \hspace{1cm}   {\rm for}  \hspace{1cm} n=1,
   2,...
\end {equation}
\begin {equation}
  E_{TOT} = \sum_{n=1} <N>_{n}E_{n} \hspace{1cm}   {\rm for}  \hspace{1cm} n=1,
  2,...
\end {equation}
It can be again supposed that expressions (23)-(29) are satisfied.

For a macroscopic black hole, when conditions (30), (31) are
satisfied, left hand of (29) represents a large number. On the
right hand of (29) terms corresponding to large n must be
approximately neglected since expression (32), according to (17),
(46), is again satisfied. But for small $n$, i.e. $n \sim 1$,
terms on the right hand of (29) are small (smaller than 1) too.
For this reason condition (29) can not be satisfied for
macroscopic fermionic black hole.

All this implies that suggested simple model of the black hole
generally cannot be applied on the fermionic black hole.

So, it can be concluded that one horizon, microscopic, bosonic
black hole behaves (partially) as a "laser". Obviously, it is in a
significant conceptual analogy with Corley-Jacobson theory [1],
[2] of the two horizon, bosonic black hole "laser".

\section {A simple solution of the information loss paradox}

In distinction from spontaneous emission of the radiation, that is
decoherent, stimulated emission of the radiation is coherent. It
represents the general characteristics of the quantum theory and
refers on the single horizon black hole too. Since by its Hawking
evaporation any macroscopic black hole turns out in a microscopic
black hole and since microscopic black hole yields partially
coherent stimulated emission of the radiation it implies, as it
will be now shown, that by total black hole evaporation there is
no decoherence, i.e. information loss.

Information loss paradox [14]-[17], representing one of the most
mysterious, unsolved to this day physical paradox, can be simply
formulated in the following way.

Suppose that there is a gravitationally collapsing, macroscopic
(with mass larger than MP) physical system that, before
gravitational collapse, is described by a pure quantum state
$|\Psi_{in}>$.

Suppose that gravitational dynamics including collapse can be
described by usual, unitary quantum dynamical evolution during
time, $\ hat {U}$, so that pure quantum state of given system
after collapse and before total evaporation caused by Hawking
radiation can be presented in the form of a super-systemic
superposition, i.e. correlated quantum state of the quantum
super-system $IN+SP$
\begin {equation}
     |\Psi^{IN+SP}> = \hat {U}|\Psi_{in}> =  \sum_{n=0}c_{n}|\Psi^{IN}_{n}> \otimes |\Psi^{SP}_{n} >    .
\end {equation}
Here $\otimes$ represents the tensorial product and $c_{n}$   for
$n=0, 1, 2,...$ . Also, ${\it B}_{IN}=  |\Psi^{IN}_{n}>$, for
$n=0, 1, 2,...$ represents a basis corresponding to sub-system
inside horizon, $IN$, while ${\it B}_{SP}=  |\Psi^{SP}_{n} >$, for
$n=0, 1, 2,...$ represents a basis corresponding to other
sub-system, i.e. spontaneous, decoherent Hawking radiation outside
horizon, $SP$.

It can be expected that Hawking thermal radiation or $SP$, can be
described by a mixed state, i.e. by the following statistical
operator
\begin {equation}
    \rho_{SP}=\sum_{n=0} | c_{n}|^{2} |\Psi^{SP}_{n} > < \Psi^{SP}_{n}| .
\end {equation}
This statistical operator can be obtained as the following
sub-systemic or second kind mixture [19] of the correlated state
(49), i.e. as the following statistical operator
\begin {equation}
        \rho_{SP}=\sum_{n=0}<\Psi^{IN}_{n}| |\Psi^{IN+SP} > <\Psi^{IN+SP}| |\Psi^{IN}_{n}> =
        \sum_{n=0} |c_{n}|^{2} |\Psi^{SP}_{n} > < \Psi^{SP}_{n}|
\end {equation}
obtained by averaging of the pure, correlated state
$|\Psi^{IN+SP}>$ (49 ) over $IN$, precisely ${\it B}_{IN}$ .

All this, exactly quantum mechanically, has the following physical
meaning [18]. Super-system $IN+SP$ is exactly completely described
by pure, correlated quantum state (49), on the one hand. On the
other hand, $SP$ as the sub-system of $IN+SP$, in respect to any
characteristics sub-systemic measurement realized outside black
hole horizon, is effectively described by mixture (51).

In this way, before time moment of the total black hole
evaporation by means of Hawking radiation, according to usual
quantum mechanical formalism, there is no any contradiction
between quantum theoretical description and expectations.

During black hole evaporation, roughly speaking, $IN$ becomes
smaller and smaller while $SP$ becomes greater and greater.
Finally, in the time moment of the total black hole evaporation
sub-system $IN$ disappear entirely. Then correlated quantum state
(49) of turns out in the final, non-correlated quantum state
\begin {equation}
       | \Psi^{IN+SP}_{fin}> = |\Psi^{IN}_{0} >\otimes \sum_{n=0} c_{n}|\Psi^{SP}_{n} >
\end {equation}
that describes two non-correlated, i.e. separated sub-systems,
totally evaporated $IN$ in pure vacuum state $|\Psi^{IN}_{0} >$
and Hawking radiation in pure state $\sum_{n=0}
c_{n}|\Psi^{SP}_{n} >$ representing a simple, sub-systemic
superposition.

Obtained pure quantum state $\sum_{n=0} c_{n}|\Psi^{SP}_{n} >$ of
SP sharply contradicts to mixed state (50), or now (50) cannot be
more presented by sub-systemic averaging of ( 52) over ${\it
B}_{IN}$ . In other word it seems that this mixture, if it really
exists, cannot be an effective, i.e. sub-systemic phenomenon. It
seems that this mixture must be a completely exact phenomenon
principally different from quantum theoretically predicted pure
state. It implies principal limitations of the quantum mechanical
descriptions. Or, it implies that by total Hawking evaporation of
a black hole there is a paradoxical, i.e. quantum mechanically
undescribable, real transition form pure in the mixed quantum
state and corresponding entropy increase or information loss.

According to many later considerations it has been concluded that,
perhaps, information loss cannot really exist in the nature since
its existence implies many non-physical (unobserved) effects, e.g.
universe heating etc. . Nevertheless, it implies other serious
problem. Namely, there is no reasonable mechanism, i.e. consistent
theory for stopping of Hawking radiation before total evaporation
of black hole (there is no completely satisfactory remnant
theory).

Now we shall suggest a simple solution of the information loss
paradox.

As it has been previously discussed black hole can emit radiation
not only spontaneously, but also stimulate. This stimulated
emission can be simply denoted by $ST$. For macroscopic black hole
stimulated emission can be approximately neglected. Then
macroscopic black hole evaporation can be satisfactorily described
by (48) including effective description of Hawking radiation by
(51).

But in the time moment when given black hole becomes microscopic
and $ST$ non-neglecting, evaporation process must be described in
the more complex way, i.e. by a more complex than (49), pure,
correlated quantum state of super-system $IN+SP+ST$
\begin {equation}
     |\Psi^{IN+SP+ST}>= \sum_{n=0}c_{n} |\Psi^{IN}_{n}> \otimes |\Psi^{SP}_{n} >\otimes |\Psi^{ST}_{n} >    .
\end {equation}
Here ${\it B}_{ST}= { |\Psi^{ST}_{n}> , for n=0, 1, 2, … }$
represents a basis corresponding to an additional sub-system
outside horizon, $ST$.  It implies a consistent description of
$SP$ by more complex (in sense of the calculation procedure) than
(51) second kind mixture, i.e. statistical operator
\begin {equation}
  \rho_{SP}=\sum_{n,k=0}<\Psi^{IN}_{n}|\otimes<\Psi^{ST}_{k}| |\Psi^{IN+SP+ST}><\Psi^{IN+SP+ST}| |\Psi^{IN}_{n}>\otimes |\Psi^{ST}_{k}> =
  \sum_{n=0} | c_{n}|^{2} |\Psi^{SP}_{n} > < \Psi^{SP}_{n}|
\end {equation}
that holds final form equivalent to (51) or (50). Obviously, given
statistical operator $\rho_{SP}$ (54) is obtained by averaging of
the pure, correlated state $|\Psi^{IN+SP+ST}>$ (53) over $IN+ST$,
precisely ${\it B}_{IN}\otimes {\it B}_{ST}$ .

Finally, in the time moment of the total black hole evaporation
$IN$ disappear so that $IN+SP+ST$ becomes described by final
quantum state
\begin {equation}
     |\Psi^{IN+SP+ST}_{fin}> = |\Psi^{IN}_{0}>\otimes \sum_{n=0} c_{n} |\Psi^{SP}_{n} >\otimes |\Psi^{ST}_{n}>    .
\end {equation}
This state describes totally evaporated IN described by pure
vacuum state $|\Psi^{IN}_{0}>$ and separated from SP+ST described
by the pure, correlated quantum state
\begin {equation}
     |\Psi^{SP+ST}_{fin}> = \sum_{n=0} c_{n} |\Psi^{SP}_{n} >\otimes |\Psi^{ST}_{n}> .
\end {equation}
It admits that SP be described again by second kind mixture (54),
or, simply by second kind mixture
\begin {equation}
  \rho_{SP}=\sum_{n=0} <\Psi^{ST}_{n}| |\Psi^{SP+ST}_{fin}><\Psi^{SP+ST}_{fin}| |\Psi^{ST}_{n}> =
  \sum_{n=0} | c_{n}|^{2} |\Psi^{SP}_{n} > < \Psi^{SP}_{n}|
\end {equation}
both of which have the same final form equivalent to (51) or (50).

All this, exactly quantum mechanically, has the following physical
meaning. Super-system $IN+SP+ST$ or, simplifying, $SP+ST$, is,
even after black hole, i.e. $IN$ total evaporation, exactly
completely described by pure, correlated quantum state (63) or
(64), on the one hand. On the other hand, $SP$ as a sub-system of
the super-system, in respect to any characteristics sub-systemic
measurement realized on $SP$ is effectively described by mixture
(54 ), i.e. (57) equivalent to expected mixture (50).

In this way, total black hole evaporation by means of Hawking
radiation can be described by usual quantum mechanical formalism,
i.e. unitary quantum dynamical evolution without any contradiction
with expectations. It represents simple solution of the
information loss paradox.

\section {Mass duality as T-duality}

Previously suggested model of the black hole, according to which
large system is surrounded by (statistical ensemble of the) small
system on the horizon surface area, in some degree similar to
atomic nucleus by electronic shell in Bohr atomic theory, is
simple and intuitively clear. But this intuitive clearness,
strictly speaking, is satisfied only in case of the macroscopic
black hole, i.e. when condition (30) is satisfied.

In opposite case, i.e. for microscopic black hole when condition
(39) is satisfied, suggested model is not intuitively clear.
Really, according to (39), mass of the small system becomes
greater than mass of the large system. But even in this case
application of the terms obtained for macroscopic black hole
yields, quite unexpectedly and contra-intuitively, correct,
non-trivial results. According to usual intuition it would be
expected that for microscopic black hole large system surrounds
small or that thermodynamical state of the statistical ensemble
changes slower than thermodynamical state of the reservoir.

We shall suppose that presented non-intuitive character of the
suggested model for microscopic black hole does not represent a
consequence of the model weakness. On the contrary, we shall
suppose that presented non-intuitive character of the model for
microscopic black hole represents the consequence of the mass
duality.

Namely, according to (6), expression
\begin {equation}
  m_{1} = \frac {1}{4\pi}\frac {M^{2}_{P}}{M}
\end {equation}
that can be considered as determination of m1 by M, and expression
\begin {equation}
  M = \frac {1}{4\pi}\frac {M^{2}_{P}}{m_{1}}
\end {equation}
that can be considered as determination of $M$ by $m_{1}$ have
analogous form. Or we can say that (58) and (59) are mutually dual
in sense that (58) can be changed by (59) and vice versa by
changing of $m_{1}$  by $M$ and vice versa.

Moreover, left hand of the condition (2) for $n=1$, according to
(1), can be transformed in the following way
\begin {equation}
  m_{1}c R = m_{1} \frac {2GM}{c^{2}} = Mc\frac {2Gm_{1}}{c^{2}}= Mc R_{1}
\end {equation}
where
\begin {equation}
   R_{1}= \frac {2Gm_{1}}{c^{2}}
\end {equation}
represents Schwarzschild radius for $m_{1}$.

According to (60), condition (2) for $n=1$, can be transformed in
\begin {equation}
   Mc R_{1} = \frac {\hbar}{2\pi}   .
\end {equation}
It, after multiplication with $n$, for $n=1, 2, …$ , yields
\begin {equation}
   M_{n}c R_{1} = n \frac {\hbar}{2\pi} \hspace{1cm}   {\rm for}  \hspace{1cm} n=1,
   2,...
\end {equation}
where
\begin {equation}
   M_{n} = n M   \hspace{1cm}   {\rm for}  \hspace{1cm}  n=1, 2, …  .
\end {equation}

Also, from (63) it follows
\begin {equation}
    2\pi R_{1} = n\frac {\hbar}{M_{n}c} = n \Lambda_{rn} \hspace{1cm}   {\rm for}  \hspace{1cm}  n = 1,
    2,...
\end {equation}
where
\begin {equation}
      \Lambda_{rn}= \frac {\hbar}{M_{n}c} \hspace{1cm}   {\rm for}  \hspace{1cm}  n = 1,
      2,...
\end {equation}
represent reduced Compton wave lengths corresponding to $M_{n}$
for $n=1, 2, …$  . Given reduced Compton wave lengths we shall
entitled dual reduced Compton wave lengths.

All this admits that quantum system with mass m1 can be treated as
a black hole, dual to black hole corresponding to $M= M_{1}$, so
that this dual black hole holds a dual large system $DL$ with mass
$m_{1}$ and dual small system $ds$ with masses $M_{n}$ for $n=1,
2, …$  .

     Then, expression (65) simply means that {\it circumference of the dual black hole horizon holds exactly} $n$ {\it corresponding} $n$-{\it th dual reduced Compton wave lengths of the dual small quantum system with mass} $M_{n}${\it  captured at the dual black hole horizon surface}, for $n = 1, 2, …$ .

Multiplication (1) and (61), according to (58), yields
\begin {equation}
  R R_{1} = \frac {1}{\pi}L^{2}_{P}
\end {equation}
where $L_{P}$ represents Planck length. It implies
\begin {equation}
  R = \frac {1}{\pi}\frac { L^{2}_{P}}{R_{1}}
\end {equation}
and
\begin {equation}
   R_{1} = \frac {1}{\pi}\frac { L^{2}_{P}}{R}
\end {equation}
that represents a duality between horizon radius of the black hole
and horizon radius of the dual black hole.

Obviously given duality between horizon radiuses is analogous to
T-duality characteristic for the string theories.

It implies that our simple, "macroscopic", i.e. phenomenological
model of the black hole can be generalized within a more accurate,
"microscopic", i.e. string theory, especially Copeland-Lahiri
theory [13]. But detailed analysis of the correspondence between
our and a string theory of the black hole goes over basic
intention of this work.

\section {Conclusion}

In conclusion we can shortly repeat and point out the following.
In this work we showed that single horizon black hole behaves as a
"laser". It is in many aspects conceptually analogous to
remarkable Corley and Jacobson work on the two horizon black hole
laser. We started by proposition that circumference of the black
hole horizon holds the natural (integer) quantum number of
corresponding reduced Compton's wave length of some bosonic
systems in great canonical ensemble. For macroscopic black hole
ground state is practically totally occupied while other states
are practically totally unoccupied which represents a typical Bose
condensation. Number of the systems in this condensate represents
black hole entropy. For microscopic black hole few lowest energy
levels are occupied with almost equivalent population (with
negative chemical potential) while all other energy states (with
positive chemical potential) are practically unoccupied. It
implies that now not only spontaneous but also stimulated emission
of radiation comparable with spontaneous emission occurs. Since by
its Hawking evaporation any macroscopic black hole turns in a
microscopic black hole and since microscopic black hole yields in
a significant degree coherent stimulated emission of the
radiation, it implies that by total black hole evaporation there
is no decoherence, i.e. information loss. Finally, a mass duality
characteristic for suggested black hole model corresponding to
string T-duality is discussed.

\vspace {1.5cm} Authors are very grateful to Prof. Dr. Petar
Gruji\'c, Prof. Dr. Darko Kapor and Prof. Dr. Miodrag Krmar for
many (infinite) different forms of the support and help. \vspace
{1.5cm}

\section {References}

\begin {itemize}

\item [[1]] S. Corley, T. Jacobson, {\it Black hole lasers}, hep-th/9806203
\item [[2]] U. Leonhardt, T. G. Philbin, {\it Black Hole Lasers Revised}, gr-qc/0803.0669
\item [[3]] V.Pankovic, M.Predojevic, P.Grujic, {\it A Bohr's Semiclassical Model of the Black Hole Thermodynamics}, Serb. Astron. J., {\bf 176}, (2008), 15; gr-qc/0709.1812
\item [[4]] V. Pankovic, J. Ivanovic, M.Predojevic, A.-M. Radakovic, {\it The Simplest Determination of the Thermodynamical Characteristics of Schwarzschild, Kerr and Reisner-Nordström black hole}, gr-qc/0803.0620
\item [[5]] V. Pankovi\'c, Simo Ciganovi\'c, Rade Glavatovi\'c, {\it The Simplest Determination of the Thermodynamical Characteristics of Kerr-Newman Black Hole}, gr-qc/0804.2327
\item [[6]] J. D. Bekenstein, Phys. Rev., {\bf D7}, (1973), 2333
\item [[7]] S. W. Hawking, Comm. Math. Phys., {\bf 43}, (1975), 199
\item [[8]] S. W. Hawking, Phys. Rev., {\bf D14}, (1976), 2460
\item [[9]] S. W. Hawking, in {\it General Relativity, an Einstein Centenary Survey}, eds. S. W. Hawking, W. Israel (Cambridge University Press, Cambridge, UK 1979)
\item [[10]] R. M. Wald, {\it Black Hole and Thermodynamics}, gr-qc/9702022
\item [[11]] R. M. Wald, {\it The Thermodynamics of Black Holes}, gr-qc/9912119
\item [[12]] D. N. Page, {\it Hawking Radiation and Black Hole Thermodynamics}, hep-th/0409024
\item [[13]] E. J. Copeland, A.Lahiri, Class. Quant. Grav., {\bf 12} (1995) L113 ; gr-qc/9508031
\item [[14]] S. W. Hawking, Phys. Rev. {\bf D 14}, (1976), 2460
\item [[15]] J. Preskill, {\it Do Black Holes Destroy Information?}, hep-th/9209058
\item [[16]] S. B. Giddings, {\it Black Hole Information, Unitarity and Nonlocallity}, hep-th/0605196
\item [[18]] S. D. Mathur, {\it What Exactly is Information Paradox?}, hep-th/0803.2030
\item [[19]] B.d'Espagnat, {\it Conceptual Foundations of the Quantum Mechanics} (Benjamin, London-Amsterdam-New York, 1976)

\end{itemize}

\end{document}